\documentclass[floatfix,showpacs,amsmath,amssymb,letterpaper,groupaddresses,superscriptaddress]{article}
\usepackage{latexsym}

\usepackage{authblk}

\usepackage{epsfig}
\usepackage{easybmat} %package for diagonal matrix form
\usepackage{arydshln}
\usepackage{amsmath}
\usepackage{slashed}
\usepackage{amssymb}
\usepackage{graphicx}
\usepackage{times}
\usepackage{bm}
\usepackage[section]{placeins}
\usepackage{array}
\usepackage{cite}
\usepackage{physics}
\usepackage[normalem]{ulem}

\usepackage{tikz}
\usetikzlibrary{matrix,positioning,decorations.pathreplacing}

\usepackage{color}

\def\a{\alpha}

\def\ni{\noindent}
\def\m{\mu}

\def\be{\begin{equation}}
	\def\ee{\end{equation}}
\def\ba{\begin{eqnarray}}
	\def\ea{\end{eqnarray}}
\def\la{\langle}
\def\ra{\rangle}
\def\a{\alpha}

\def\m{\mu}

\def\h{\hskip 1cm}

\def\lo{\longrightarrow}
\def\bx{{\bf x}}
\def\by{{\bf y}}

\title{\Large \bf Quantum coherence between subspaces: State transformation, Cohering Power, $k$-coherence and other properties}

\author[1]{A. Mani}
\affil[1]{Department of Engineering Science, College of Engineering, University of Tehran, Iran}

\author[2]{F. Rezazadeh}
\author[2]{V. Karimipour}
\affil[2]{Department of Physics, Sharif University of Technology, Tehran, Iran}

\begin{document}
	
	\maketitle

\begin{abstract}

	The concept of bock-coherence, first introduced in \cite{Aberg} and developed in \cite{Bischof1,Bischof2}  encompasses the case where experimental capabilities are not so delicate  to perform arbitrary refined measurements on  individual atoms.  We  develop a framework which facilitates further investigation of this resource theory in several respects. Using this framework, we investigate the
	problem of state conversion by incoherent operations and show that a majorization condition is the necessary and sufficient condition for state transformation by block-incoherent operations.  We also determine the form of the maximally coherent state from which all other states and all unitary gates can be constructed by incoherent operations.  Thereafter, we define the concept of block-cohering and block-decohering powers of quantum channels and determine these powers for several types of channels. Finally, we explore the relation between block coherence and a previous extension of coherence, known as $k$-coherence. \\

\end{abstract}

	
\vskip 2cm

\section{Introduction}
%\textcolor{blue}{and has lots of applications in quantum technologies \cite{coherence application1,coherence application2,coherence application3,coherence application4}}

From the very beginning of quantum theory, coherence of states, as a property which radically distinguishes the quantum superposition from classical mixtures, has been the subject of much discussion.  While coherence has not been entirely unfamiliar to physicists, as it has been present in all forms of wave phenomena, only in quantum mechanics it has revealed its most exciting properties. It is here that one encounters intriguing concepts like superposition of spacial degrees of freedom as in double-slit experiment \cite{marcus1,marcus2}, superposition of macroscopic states of many body systems \cite{fazio1, tahereh} as in quantum phase transition, and entanglement which itself underlies the unique features of quantum computation and quantum information processing \cite{nielson}.  A plethora of theoretical and experimental techniques have been known for manipulating coherence in optical experiments \cite{Glauber, Sudarshan, Mandel}, and theoretical limitations for manipulation of superposition has been studied in various works \cite{sup1,sup2}. Nevertheless, attempts for quantifying superposition of orthogonal states, are rather recent. General measures of coherence were first introduced by Aberg in \cite{Aberg} and then formulated in a quantitative resource-theory based form in \cite{Plenio, Levi} which was further developed in various directions in  
\cite{ Xu, Plenio2, Suman, Marvian, Streltsov1, Streltsov2, Winter, Gour1, Gour2, Mani, Zanardi, Yadin, Bromley, Adesso, Kim, Bischof1, Bischof2, blockcoherence measure, Kcoherence1, Kcoherence2,Kcoherence3,Kcoherence4}. \\

\ni The resource theory was itself inspired by the understanding that entanglement can be considered as a resource which is used in an efficient and useful way and consumed at the end of most quantum communication tasks \cite{resource theory, entresource1, entresource2,entresource3}. Likewise, superposition and coherence, can also be thought of as a kind of resource which is used in a quantum process and consumed at the end. The core concept of any resource theory is the operational restrictions that we have for manipulating quantum states in our laboratory \cite{assymetryResource1, assymetryResource2, assymetryResource3, resource1, resource2, resource3, transform}. In entanglement these restrictions derive from locality, and in superposition and coherence, they derive from our insufficient means for accessing any kind of basis for quantum states, either in our measurements and filtering operations or in other kinds of operations. For example a laboratory may easily put spin one-half particles in states $|\uparrow, z\ra$ or $|\downarrow, z\ra$, but not in their arbitrary superposition. To summarize the notions introduced in \cite{Plenio}, a basis $\{|i\ra, i=1\cdots d\}$ for a Hilbert space is chosen as the preferred basis. A state is called incoherent, if it is diagonal in this preferred basis, i.e. if 
\be \label{standardINC}
\rho_{inc} = \sum_{i=0}^{d-1} p_i |i\ra \la i|,
\ee
where $\{p_i\}$ is a probability distribution. In fact, the incoherent states are the ones which can be freely generated by the measurements of experimenter in the preferred basis. 
The totality of such states form a convex set ${\cal I}_{inc}$ in the space of all conceivable quantum states. \\
	The incoherent operations are then defined to be the ones which do not generate any coherence from an incoherent state, i.e. they are trace preserving completely positive operations which map the set of incoherent states into itself, i.e. a quantum operation ${\cal E}$ is incoherent if ${\cal E}: {\cal I}_{inc}\lo {\cal I}_{inc}.$ \\
	 Finally a Maximally coherent state is a state from which all other states and all unitary operators (quantum gates), can be constructed purely by incoherent operations, i.e. by operations that are at disposal of the experimenter in his or her laboratory.  
It was shown in \cite{Plenio} that a state like 
\be
|\psi \ra = \frac{1}{\sqrt{d}}\sum_{i=1}^{d} |i\ra,
\ee
is a  maximally coherent state of a $d$-dimensional Hilbert space, in the above sense. Therefore it is a resource state in the context of coherence theory. 
Once the free states, and free operations were recognized, quantitative measures of coherence could be defined in the spirit of resource theory. \\

\ni This primary resource theory of coherence is based on an orthogonal basis for description of the density matrix and rank-one measurement of the experimenter, and it leads to measures of coherence which should be expressed in terms of the individual matrix elements of the density matrix, the determination of which may not be experimentally feasible. The removal of the constraints of this standard resource theory of coherence, has led to other generalized theories. For example, the requirement of orthogonality of the basis vectors is relaxed to their linear independence  in \cite{Plenio2}, and also the authors of \cite{Suman} write the density matrix in terms of expectation values of Hermitian operators and express the known measures of coherence in terms of what they call the observable matrix, all the elements of which are directly measurable in the laboratory. \\

In a different development, the author of \cite{Aberg} has introduced the notion of block coherence and different block coherence measures are defined in \cite{Aberg,Xu, blockcoherence measure}. In the resource theory of block coherence, the block-incoherent states have a block diagonal structure which is determined by a projective (not necessarily rank one) measurement. The resource theory of coherence based on positive-operator-valued measurements (POVM) is also introduced in \cite{Bischof1}, where the authors use the Naimark theorem to define the POVM-based coherence. This generalization is also quantified in \cite{ blockcoherence measure,Xu, Bischof2}.
 \\

\ni The theory of block coherence is of special importance in cases where the experimenter does not have an ability to  measure a complete set of observables and prepare a complete basis of states, which is often the case. Mathematically this means that the projectors of measurement are not rank-one projectors. For example one  may only be interested in measuring a property of a group of particles, in which case the projectors will be $\Pi_j=\mathbb{I}\otimes \mathbb{I}\cdots \otimes\pi_j\otimes\cdots \mathbb{I}\otimes \mathbb{I}$, where $\pi_j$ are projectors on that specified group. This is the case where the projectors $\Pi_j$ are no longer of unit rank.  Even for one particle, one may only be able to  measure its total spin and not the $z$-component of its spin. Or one may only be able to determine whether the spins of two particles are parallel or anti-parallel, e.g. in a communication task where these pairs of  particles are sent between two parties with no shared reference frame \cite{Gisin, Massar, Bartlett1, Bartlett2, ours1, ours2, ours3, ours4}. In other quantum protocols, one may need to determine whether the majority of spins are up or down in a given 
precision \cite{hetero}. All these refer to realistic situations where a preferred and complete basis and the refined operations induced by that measurement are not accessible for us. Under such circumstances, we should adapt our notions and measures of coherence to these new limitations. For example, in a situation where we can only do projective measurements with $$\pi_1=|0\ra\la 0|+|1\ra\la 1|\ \ \ , \pi_1=|2\ra\la 2|,$$ \\
in a 3-level system, it is meaningless to assign non-zero coherence to a state like $a|0\ra+b|1\ra$ and zero coherence to a state like $a^2|0\ra\la 0|+b^2|1\ra\la 1|$. \\

\ni Besides its theoretical interest, the resource theory of block coherence may have significant practical consequences. It is important to know that the availability of less-refined measurements in a lab, affects our definition of incoherent states and operations and the resourceful states.  Do we need more or less coherence, according to the initial definition of \cite{Plenio}, in order to produce a certain state?  Shall we need more complicated incoherent operations to construct arbitrary states from our resource states? Which states can be transformed to each other freely?  Can one define block-cohering and block-decohering powers of quantum channels as in \cite{Mani, Zanardi}? What is the relation of block-coherence and the notion of $k$-coherence developed in \cite{Kcoherence1, Kcoherence2, Kcoherence3, Kcoherence4}.  These questions have been left  unanswered in previous studies, and as we will see, these questions guide us to a rich and comprehensive structure for the resource theory of block coherence.\\

In the present work, we will provide answers to these questions. To this end, we first introduce a mathematical framework, which facilitates many of the consequent calculations. Then we prove  a majorization-like sufficient and necessary condition for pure state transformation and  find the explicit form of the incoherent operations which perform  state conversion.  We also explicitly show that one can use the action of incoherent operations on the maximally coherent sate, to construct any arbitrary gate.  We will see that, the more course-grained our measurements are,  more complicated incoherent operations are necessary to convert this state to an arbitrary state and construct an arbitrary unitary operation.  We also define the block-cohering and block-decohering powers of quantum channels, and derive  closed formulae of these quantities for several families of quantum channels. Finally we elaborate on the relation between block-coherence and an interesting  notion called $k$- coherence \cite{Kcoherence1, Kcoherence2, Kcoherence3, Kcoherence4}. The later concepts which is different from block coherence is based on the number of  basis states which are in a superposition, in a given general state. We find a curious and interesting relation between the two notions which we will clarify by an explicit and yet general example. \\

\ni The structure of the paper is as follows. In section (\ref{Notation}) we state our notations and conventions. In section (\ref{prelim}) we recapitulate the previous results in a simple mathematical form. We then briefly review two of the previously defined block coherence measures in section (\ref{measures}). In section (\ref{PureStateConversion}) we investigate the pure state conversion and show that in the context of block coherence, majorization is still a sufficient condition for state transformation by incoherent operations.  We show in section (\ref{construction}) how by having access to a maximally incoherent state and by using only incoherent operations, one can implement any arbitrary unitary gate. To this end we obtain the explicit form of the appropriate Kraus operators. We also define the block cohering and decohering powers of a quantum map in section (\ref{powers}) and we calculate these powers for a few families of channels. The relation between block coherence and $k$-coherence is investigated in section (\ref{relation}). The paper ends with a conclusion. Finally in an appendix we show that the majorization condition of section (\ref{PureStateConversion}) is also necessary for state transformation by block-incoherent operations.

\section{Notations} \label{Notation}
Let $\mathcal{H}$ be a Hilbert space which can be decomposed into subspaces such that
\be \label{Hilbert}
\mathcal{H} =\oplus_{\mu=1}^M H_\mu,
\ee with ${\rm dim}(H_\mu)=d_\mu $ and let $\pi_{\mu}$ be the projection operator on the subspace $H_\mu$:
$$\sum_{\mu=1}^M \pi_\mu=\mathbb{I}_{\mathcal{H}}.$$ These projectors define the only measurements that are at disposal in our laboratory. 
Let the subspace $H_\mu$ be spanned by an orthonormal basis $\{|e_{i_\mu}\rangle,\ i_\mu=1\cdots d_\mu\}$. Here we have abbreviated the more detailed notation $e^{(\mu)}_{i_\mu}$ where ${(\mu)}$ points to the subspace and $i_{\mu}$ to the basis state in the subspace, simply to $e_{i_\mu}$, hoping that this will not lead to confusion. Obviously we have $$\la e_{i_\mu}|e_{j_\mu}\ra=\delta_{i_\mu,j_\mu}, \h  {\rm and }  \h  \la e_{i_\mu}|e_{j_\nu}\ra=0\ \ \ 
 \ \ \ \mu\ne \nu.$$ We also define an auxiliary space $$Q={\rm Span} \{|1\ra, |2\ra, \cdots , |\mu\ra, \cdots , |M\ra\},$$ to specify different sub-spaces in the following way. A block diagonal operator is then denoted as 
\be
A=\sum_{\mu=1}^M  |\mu\rangle \la \mu|\otimes A_\mu,
\ee
and an operator which is non-zero only on the block $\mu\nu$ is written as $B= |\mu\rangle \langle \nu|\otimes B_{\mu\nu}$. Such an operator maps $H_\nu$ to $H_\mu$ and acts as zero operator on all other subspaces.

\section{Preliminaries}\label{prelim}
\ni Consider the $d$-dimensional Hilbert space $ \cal H$  in (\ref{Hilbert}), and let $\mathcal{M}=\{\pi_\mu | \mu=1 \cdots M\}$ describe a measurement with projective operators $\pi_\mu$, not necessarily of rank one. 
A quantum state is defined to be block incoherent if it is in block-diagonal form, that is if 
\be
\rho_{inc}=\sum_{\mu=1}^M p_\mu |\mu\rangle \langle \mu| \otimes \rho_\mu,\h \sum_\mu x_\mu=1,
\ee
in which each $\rho_\mu$ is a density matrix in $H_\mu$. The outcome of any measurement $\mathcal{M}$ on any state is of the above form. Incoherent states, when being measured remain intact. In more explicit form, an incoherent state has the matrix form
\be \label{incohstate}
\rho_{inc}=\begin{pmatrix}
    p_1\rho_1&&&&\\ &p_2\rho_2&&&\\ &&.&&\\&&&.&\\ &&&&p_M\rho_M
\end{pmatrix},
\ee
where $\rho_\mu$ is a $d_\mu\times d_\mu$ density matrix. When $d_\mu=1$ for all $\mu=1 \cdots M$, this definition coincides with the usual definition of incoherent states. Obviously the set of all block incoherent states is a convex set which is denoted by ${\cal I}_{inc}$. Note that in any subspace no preferred basis is assigned.\\

\ni Incoherent operations are the ones which do not create coherence out of incoherent states. Different approaches are used to define these operations \cite{Gour2}. The largest class of incoherent operations are the so-called MIO (Maximal Incoherent Operations) and consist of all operations which maps ${\cal I}_{inc}$ to itself. A quantum operation is said to be IO (Incoherent Operation), if it has a Kraus representation  such that each Kraus operator maps ${\cal I}_{inc}$ to itself.\\

\ni An operation ${\cal E}=\sum_a K_a\rho K_a^\dagger$ is Block-IO if and only if its Kraus operators have the following form \cite{Bischof2} 
\be \label{inco operation}
K_a = \sum_\mu   |a(\mu)\ra\la \mu| \otimes K^a_\mu,
\ee
in which $a: \{1, 2, \cdots,  M\}\lo \{1,2,\cdots,  M\}$ is an arbitrary function, not necessarily a permutation, and $K_\mu^a $ is any arbitrary operator. The proof is presented in \cite{Bischof2}, but can also be shown straightforwardly by using the auxiliary space introduced in section (\ref{Notation}), one simply notes that a Kraus operator of the form $K_a = \sum_\mu   |a(\mu)\ra\la \mu| \otimes K^a_\mu$, when acting on 
$ \rho =\sum_{\mu} p_\mu  |\mu\rangle \langle \mu| \otimes \rho_\mu $, produces another incoherent state of the same form. \\

\ni The explicit form of  these Kraus operators are such that in each column only one block should be non-zero. For example, when $M=2$, regardless of the dimensions of blocks, the admissible form of Kraus operators are as follows:  
\be\label{exxx}
K_1=\begin{pmatrix}
    A_1&\\ &A_2
\end{pmatrix},\ \ \
K_2=\begin{pmatrix}
    B_1&B_2\\ &
\end{pmatrix},\ 
K_3=\begin{pmatrix}
    &C_1\\ C_2&
\end{pmatrix},\ \ \
K_4=\begin{pmatrix}
    &\\ D_1&D_2
\end{pmatrix}.
\ee

\section{Measures of Block Coherence}\label{measures}
\ni{\bf 1- The measure based on relative entropy:}
Given an arbitrary state $\rho$, one can define its measure of block-coherence as its minimum distance from the set of block-incoherent states. This approach has been followed in \cite{Aberg, Xu, blockcoherence measure, Bischof1, Bischof2}, which lead to the following  block coherence content of the state $\rho$ is 
\be \label{distance measure}
C(\rho)=\min_{\delta\in {\cal I}_{inc}} D(\rho, \delta),
\ee	
where $D$ is any distance. Although relative entropy does not have all the properties of distance, it is usually used in measures like (\ref{distance measure}) to quantify various resources. If one takes $D(\rho,\delta)$ to be the relative entropy between the two states, then the closest block incoherent state to a given state $\rho$ is obtained by simply removing all the off-diagonal blocks in the density matrix \cite{Aberg}.	Hence a closed and easily calculable formula for the entropy based measure of block coherence of a given state $\rho$ is
\be \label{closedform}
C^M_s(\rho)=S(\rho^*)-S(\rho).
\ee
where 
\be
\rho^*:=\sum_{\mu=1}^M \pi_\mu \rho \pi_\mu = \sum_{\mu=1}^M  |\mu\ra\la \mu|\otimes \rho_\mu,
\ee
We now ask what kind of pure state has the largest value of block coherence, when we fix the measurement or the block structure. Consider an arbitrary pure state, 
\be\label{general}|\Psi\ra=\begin{pmatrix}x_1 |\psi_1\ra\\ x_2 |\psi_2\ra\\ \cdot\\ \cdot\\ x_M |\psi_M\ra\end{pmatrix}, \ee
 subject to $\sum_{\mu=1}^M |x_\mu |^2=1$ where $|\psi_\mu\ra$'s are arbitrary normalized pure states of dimension $d_\mu$. Using (\ref{closedform}), we find that 
\be
C^M_s(|\Psi\ra)=S(\rho^*)=-\sum_\mu |x_\mu |^2 \log |x_\mu |^2,
\ee
which means that the highest coherence belongs to states of the form 
\be\label{MC1}
|\Psi\ra_{MC}=\frac{1}{\sqrt{M}}\begin{pmatrix}|\psi_1\ra\\ |\psi_2\ra\\ \cdot\\  \cdot\\ |\psi_M\ra\end{pmatrix}, \ee
\ni i.e. the highest coherence corresponds to the states with equal probabilities of the subspaces. It is important to note that the states $|\psi_\mu\ra$ are arbitrary as long as they are not zero. They need not have any coherence in their own subspace at all.\\

\ni {\bf 2- The measure based on the $l_1 $-norm:}
Starting from $\rho^*$, another measure of block coherence based on the $l_1$-norm, which is a natural generalization of this measure for the coherence introduced in \cite{Plenio}, is defined as \cite{Xu}:
\be \label{l1norm}
C^M_{1}(\rho)=\Vert \rho - \rho^* \Vert_{l_1}=\sum_{\mu\ne \nu}\Vert \rho_{\mu\nu}\Vert_{1},
\ee
where $\rho_{\mu\nu}$ is the matrix in the block $\mu\nu$ of $\rho$ and 
\be
\Vert A\Vert_{1}=Tr(\sqrt{A^\dagger A}),
\ee
is the trace-norm of $A$.  For this measure, the block coherence of a general pure state (\ref{general}) is found to be
\be
C_1^M(|\Psi\ra)=\sum_{\mu\ne \nu} |x_\mu x_\nu|\  \Vert|\psi_\mu\ra\la \psi_\nu|\Vert
\ee
and since $\Vert\ |\psi\ra\la \phi|\ \Vert=\sqrt{\la \psi|\psi\ra \la\phi|\phi\ra}$, 
\be\label{newcoherence}
C_1^M(|\Psi\ra)=\sum_{\mu\ne \nu} |x_\mu x_\nu|.
\ee
According to this measure, the maximally coherent state (\ref{MC1}) has a coherence given by
\be\label{MaxCoh}
C_1^M(|\Psi\ra_{MC})=M-1. 
\ee

\ni Obviously, when there is only one block, there is no coherence and when all blocks are one dimensional, ($M=d$) this measure coincides with the usual measure of coherence \cite{Plenio}.\\

\ni It should be note that, by using both measures (\ref{distance measure}) and (\ref{l1norm}),  the arbitrary state (\ref{general}) has the same coherence as the state 
\be\label{general2}
|\Phi\ra=\begin{pmatrix} x_1 |\phi_1\ra\\ x_2 |\phi_2\ra\\ \cdot\\  \cdot\\ x_M |\phi_M\ra\end{pmatrix}, \ee
where
\be\label{phimu}
|\phi_\mu\rangle = \frac{1}{\sqrt{d_\mu}}\sum_{i_\mu=1}^{d_\mu}|e_{i_\mu}\ra,
\ee
is the maximally coherent state of the subspace $H_\mu$. In fact the states (\ref{general}) and (\ref{general2}) are equivalent since they can be converted to each other by applying block diagonal incoherent unitary operators of the form $U=\oplus_{\mu=1}^{M} u_\mu$. For future uses, we also state that the maximally coherent state (\ref{MC1}) is equivalent to the state
\be\label{MC2}
|\Phi\ra_{MC}=\frac{1}{\sqrt{M}}\begin{pmatrix}|\phi_1\ra\\ |\phi_2\ra\\ \cdot\\  \cdot\\ |\phi_M\ra\end{pmatrix}, \ee
with regard to their coherence. 
\\

\ni For the maximally coherent states, by suppressing the notation for states, we have, 
\be
0=C^1_{l_1}\leq C^2_{l_1}\leq \cdots C^d_{l_1}=d-1,
\ee
which shows that the value of coherence increases as $M$ increases from $1$ to $d$, a result which is expected on physical grounds, and for $M=d$ we find
$C^d_{l_1}(|\Phi\ra)=d-1$
which understandably coincides with the $l_1$ value of standard definition of coherence.
In section (\ref{PureStateConversion}) we will show that for any value of $M$ and for both types of coherence measures (\ref{distance measure}) and (\ref{l1norm}), it is indeed possible to start from the maximally coherent state for that partition, and obtain any other arbitrary state by simply using the incoherent operations, allowable for that type of partition. This verifies that these two measure are indeed  correct measures of block coherence in terms of resource theory.\\  

\ni {\bf Remarks:} \\

\ni 1- Both measures of coherence are invariant under multiplication of each state $|e_{i_\mu}\ra$ by a local phase $e^{i\phi_{i_\mu}}|e_{i_\mu}\ra.$ Therefore, we take all the coefficients in maximally coherent states to be real. \\

\ni 2- Throughout the text, we use $|\Psi\ra$ and $|\psi\ra$ to show arbitrary states the form (\ref{general}), and we use the notation $|\Phi\ra$ and $|\phi\ra$ to denote the states of the form (\ref{general2}) and (\ref{phimu}).
 \\

\section{Pure state conversion by block-incoherent operations }\label{PureStateConversion}
In this section, by explicit analytical derivation of Kraus operators, we show that in the context of block coherence, majorization is  the sufficient condition for pure state conversion by incoherent operations (IO). In the appendix, we will show that it is also the necessary condition.  Before proceeding let us remind the definition of majorization. Consider two probability distributions ${\bf p}=(p_1\geq p_2\geq \cdots \geq p_M)$ and ${\bf q}=(q_1\geq q_2\geq \cdots \geq q_M)$. We say that ${\bf p}$ majorizes ${\bf q}$ and write ${\bf p}\succ {\bf q}$ if for all $k=1,2,\cdots M$, it holds that $\sum_{i=1}^k p_i\geq  \sum_{i=1}^k q_i$.
To avoid cluttering of notations, we first explain the idea by considering the case where there are two subspaces of arbitrary dimensions and then we will then extend the argumen to the general case, where there are  arbitrary number of subspaces.

\subsection{The case where there are two subspaces of arbitrary dimensions}
Consider the case where we have only two subspaces of dimensions $d_1$ and $d_2$, i.e. $\mathcal{H}=H_1\oplus H_2$, with corresponding projectors $$\pi_1=\sum_{i=1}^{d_1}|e_i\rangle\langle e_i|\ \ \ \ {\rm and }\ \  \ \  \pi_2=\sum_{j=1}^{d_2}|f_j\rangle \langle f_j|.$$  Our task is to show that the initial state \begin{equation}\label{general-2-ini}
	|\Psi_{\bf x}\ra = \begin{pmatrix} x_1 |\psi_1\rangle \\ x_2 |\psi_2\rangle
	\end{pmatrix},\h x_1^2+x_2^2=1, 
\end{equation}
can be converted by IO to the final state
\begin{equation}\label{general-2-fin}
	|\Psi_{\bf y}\ra = \begin{pmatrix} y_1 |\psi'_1\rangle \\ y_2 |\psi'_2\rangle
	\end{pmatrix},\h y_1^2+y_2^2=1,
\end{equation}
where each of  $|\psi_i\rangle $ and $|\psi'_i\ra$ are arbitrary normalized states in their own subspaces, if the majorization condition is valid for the probability vectors  ${\bf{x}}=(x_1^2, x_2^2)$ and ${\bf{y}}=(y_1^2,y_2^2)$, i.e. ${\bf{y}}\succ {\bf{x}}$. Note that we have taken the coefficients $x_1$, $x_2$, $y_1$ and $y_2$ to be real, since block-diagonal unitary operators can always remove any phases from these numbers.\\

\noindent First of all we use the fact that in the context of block coherence, the block unitaries of the form $U_1 \oplus U_2$ are regarded as free incoherent operations, hence the states (\ref{general-2-ini}) and (\ref{general-2-fin}) can freely be converted to 
\begin{equation}\label{2-ini}
	|\Phi_{\bf x}\ra = \begin{pmatrix} x_1 |\phi_1\rangle \\ x_2 |\phi_2\rangle
	\end{pmatrix},\h x_1^2+x_2^2=1,
\end{equation}
and 
\begin{equation}\label{2-fin}
	|\Phi_{\bf y}\ra = \begin{pmatrix} y_1 |\phi_1\rangle \\ y_2 |\phi_2\rangle
	\end{pmatrix},\h y_1^2+y_2^2=1,
\end{equation}
respectively, where
\be
|\phi_1\rangle =\frac{1}{\sqrt{d_1}}\sum_{i=1}^{d_1}|e_i\rangle,  \ \ {\rm and } \  \  \ |\phi_2\rangle =\frac{1}{\sqrt{d_2}}\sum_{i=1}^{d_2}|f_i\rangle. 
\ee
are the maximally coherent states of their own subspaces, and to study the state conversion problem it will be enough to investigate the conversion from (\ref{2-ini}) to (\ref{2-fin}), without loss of generality. \\

\noindent Now consider the following generalized incoherent Kraus operators (in block-diagonal) and (anti-block-diagonal) form 
\be
A_{0}=\gamma_0 \begin{pmatrix}
	\frac{y_1}{x_1} \mathbb{I}_{d_1} &\\ & \frac{y_2}{x_2} \mathbb{I}_{d_2} \\
\end{pmatrix},
\ee
and 
\be
A_{ij}=\gamma \begin{pmatrix}
	& \frac{y_1 } {x_2 \sqrt{d_1}}|\phi_1\rangle \langle f_j|\\   \frac{y_2 }{x_1 \sqrt{d_2}}|\phi_2\rangle \langle e_i| &\\
\end{pmatrix},
\ee
where the coefficients $\gamma_0$ and $\gamma$ are considered to be real without loss of generality. The Completely Positive Trace preserving map ${\cal E}_{inc}$ can then be defined as 
\be
{\cal E}_{inc}(\rho)=A_0 \rho A_0^\dagger +\sum_{ij}A_{ij}\rho  A^\dagger_{ij}.
\ee

\noindent It is now straightforward to check that 
\be
A_0 |\Phi_{\bf x}\ra=\gamma_0 |\Phi_{\bf y}\rangle \ \ \ \ {\rm and }\ \ \ \  A_{ij}|\Phi_{\bf x}\ra=\frac{\gamma}{\sqrt{d_1 d_2}} |\Phi_{\bf y}\rangle 
\ee
which lead to 
\be
{\cal E}_{inc}(|\Phi_{\bf x}\ra\la \Phi_{\bf x})=|\Phi_{\bf y}\rangle \langle \Phi_{\bf y}|.
\ee
Note that we have used the trace-preserving property
\be
A_0^\dagger A_0+\sum_{ij}A_{ij}^\dagger A_{ij}=\mathbb{I}_{d_1+d_2},
\ee
which leaves the following constrains on the coefficients $\gamma_0$ and $\gamma$
\begin{equation}\label{maj2}
	\gamma_0^2 (\frac{y_1}{x_1})^2  + \gamma^2 (\frac{y_2}{x_1})^2  =1  , \ \ \ \ {\rm and }\ \ \ \  \gamma_0^2 (\frac{y_2}{x_2})^2  + \gamma^2 (\frac{y_1}{x_2})^2  =1.
\end{equation}
One can now easily check that the above conditions (i.e. positivity of $\gamma_0^2$ and $\gamma^2$) can be satisfied if and only if ${\bf{x}}\prec {\bf{y}}$. To see this, multiply both equations of (\ref{maj2}) to $(x_1 x_2)^2$ and then subtract them from each other, which after simplification leads to 
\be
\gamma_0^2(x_2^2  y_1^2 -x_1^2 y_2^2 ) +\gamma^2(x_2^2 y_2^2-x_1^2 y_1^2)=0.
\ee
Now suppose that $x_2 <x_1$ and $y_2<y_1$, then positivity of $\gamma_0^2$ and $\gamma^2$ together with the normalization of probability vectors ${\bf{x}}=(x_1^2, x_2^2)$ and ${\bf{y}}=(y_1^2,y_2^2)$, imply that $x_1<y_1$.  Having the same arguments for the other possible orderings of $\{x_1,x_2\}$ and $\{y_1,y_2\}$, we see that conditions (\ref{maj2}) are equivalent to the majorization condition  ${\bf x}\prec {\bf y}$. Here it should be emphasized that the dimensions $d_1$ and $d_2$ of the subspaces are not necessarily equal, and the derived majorization condition is solely based on the coefficients $x_i$ and $y_i$, regardless of the dimensions of subspaces. 
Finally, notice that the role of off-diagonal blocks in $A_{ij}$ is crucial, otherwise one cannot satisfy the trace-preserving condition necessary for the quantum operation. 

\subsection{The case where there are arbitrary number of subspaces of arbitrary dimensions}
The method of previous section can be generalized to this case in a straightforward manner. We only need to use a compact notation, as remarked in the beginning of the paper. 
With the notation introduced in section (\ref{Notation}), and following the discussion presented after equations (\ref{general-2-ini}) and (\ref{general-2-fin}), it will be enough to investigate the convertibility of the initial general state 
\be
|\Phi_\bx\ra=\sum_{\mu=1}^M x_\mu   |\mu\ra \otimes|\phi_\mu\rangle, ,\h \sum_{\mu=1}^Mx_\mu^2=1,
\ee
to the state 
\be
|\Phi_\by \ra=\sum_{\mu=1}^M y_\mu  |\mu\ra \otimes|\phi_\mu\rangle,\h \sum_{\mu=1}^My_\mu^2=1,
\ee
where  $$|\phi_\mu\ra=\frac{1}{\sqrt{d_\mu}}\sum_{i_{\mu}=1 }^{d_\mu}|e_{i_\mu}\ra,$$
was defined in (\ref{phimu}). Without loss of generality, we assume that all the coefficients are positive. This assumption is justified because block unitary operators can remove any phase from these coefficients. 
\\
\noindent Let us define the block diagonal operator $A_0$ as
\be \label{A0}
A_0:=\gamma_0 \sum_{\mu=1}^M \frac{y_\mu}{x_\mu} \  |\mu\ra\la \mu|\otimes \mathbb{I}_{d_\mu} ,
\ee
which in matrix form looks like
\be
A_0=\gamma_0 \begin{pmatrix}
	\frac{y_1}{x_1} \mathbb{I}_{d_1} &&&&\\ &\frac{y_2}{x_2}  \mathbb{I}_{d_2} &&&\\ &&.&&\\&&&.&\\ &&&&\frac{y_M}{x_M}  \mathbb{I}_{d_M} 
\end{pmatrix}.
\ee
Let $I=(i_1,i_2,\cdots i_M)$, where $i_\mu\in \{1,2,\cdots d_\mu\}$, and  set $\slashed{d}:=d_1d_2\cdots d_M$. Then for any $s\in \{1,2,\cdots M-1\}$ we define the incoherent Kraus operators
\be  \label{AIs}
{A^s}_{I}= \frac{\gamma_s}{\sqrt{\slashed{d}}} \sum_{\mu=1}^M \frac{y_\mu}{x_{\mu+s}} \sqrt{d_{\mu+s}} |\mu\ra\la \mu+s| \otimes |\phi_\mu\ra\la e_{i_{\mu+s}}|.
\ee
\noindent Direct calculation now shows that:\\

\noindent i)
\be
A_0 |\Phi_\bx\ra=\gamma_0 |\Phi_\by\ra, \h  \h
\ee
\noindent ii)
\be
A^s_{I}|\Phi_\bx\ra=\frac{\gamma_s}{\sqrt{\slashed{d}}} |\Phi_\by\ra, \h \forall s, I,
\ee
\noindent iii)
The quantum operation 
\be \label{FinalE}
{\cal E}(\rho)=A_0 \rho A^\dagger_{0}+\sum_{s,I} A^s_{I}\rho {A^s_{I}}^\dagger,
\ee
is trace-preserving if and only if 
\begin{equation} \label{major}
	\gamma_0^2  y_\mu^2  + \sum_{s=1}^{M-1} \gamma^2_s y_{\mu-s}^2 = x_\mu^2  , \h \forall \mu.
\end{equation}
\\
\ni The above statements show  that the incoherent quantum operation ${\cal E}$ can convert the state 
$|\Phi_\bx\ra$ to $|\Phi_\by\ra$, provided that the equality (\ref{major}) holds. This equality is nothing but the condition that the vector ${\bf y}=(y_1^2, y_2^2,\cdots y_M^2)$  majorizes ${\bf x}=(x_1^2,x_2^2,...,x_M^2)$ , denoted as ${\bf x}\prec {\bf y}$ \cite{nielson ,proof}.
Note that any probability vector ${\bf{y}}$ majorizes the the normalized coefficient vector $\frac{1}{M}(1,1,...,1)$ and hence any quantum state can be obtained by applying suitable incoherent operation on the maximally coherent states (\ref{MC1}), as it is expected from the resource theory of block coherence. \\

\ni It is instructive to explicitly show this last conversion by another explicit example which conveys the basic idea in a simple and yet general way.  
Let the Hilbert space be partitioned into three parts, $\mathcal{H}=H_1\oplus H_2\oplus H_3$, with dimensions $d_1, d_2$ and $d_3$ respectively. The orthonormal bases of these Hilbert spaces are respectively given by $\{|e^1_i\rangle, \ i=1\cdots d_1\}$, $\{|e^2_j\rangle,\  j=1\cdots d_2\}$ and $\{|e^3_k\rangle, \  k=1\cdots d_3\}$.  Then, the Kraus operators (\ref{A0}) and (\ref{AIs}) take the following matrix form:
\be
A_{0}=\gamma_0 \begin{pmatrix}
\frac{y_1}{x_1} \mathbb{I}_{d_1}&&\\ &\frac{y_2}{x_2} \mathbb{I}_{d_2}&\\ &&\frac{y_3}{x_3} \mathbb{I}_{d_3}
\end{pmatrix},
\ee
\begin{eqnarray}
A_{ijk}^{1}&=&\frac{\gamma_1}{\sqrt{\slashed{d}}} \begin{pmatrix}
		 0 & \frac{y_1}{x_2} \sqrt{d_2} |\phi_1\ra\la e_j^{2}| &0\\ 0 & 0 &\frac{y_2}{x_3} \sqrt{d_3} |\phi_2\ra\la e_k^{3}|\\ \frac{y_3}{x_1} \sqrt{d_1} |\phi_3\ra\la e_i^{1}| &0&0 
	\end{pmatrix} ,\cr
A_{ijk}^{2}&=&\frac{\gamma_2}{\sqrt{\slashed{d}}} \begin{pmatrix}
	0 &0& \frac{y_1}{x_3} \sqrt{d_3} |\phi_1\ra\la e_k^{3}|\\ \frac{y_2}{x_1} \sqrt{d_1} |\phi_2\ra\la e_i^{1}|& 0 & 0\\ 0& \frac{y_3}{x_2} \sqrt{d_2} |\phi_3\ra\la e_j^{2}| &0 
\end{pmatrix}.
\end{eqnarray}
It is now easy to check that
\begin{equation}
	A_0^\dagger A_0=\gamma_0^2 \begin{pmatrix}
		(\frac{y_1}{x_1})^2 &&\\ &(\frac{y_2}{x_2})^2 &\\ &&(\frac{y_3}{x_3})^2
	\end{pmatrix},
\end{equation}
and 
\begin{equation}
	\sum_{i,j,k} A_{ijk}^{1 \dagger} A_{ijk}^1=\gamma_1^2 \begin{pmatrix}
			(\frac{y_3}{x_1})^2 &&\\ &(\frac{y_1}{x_2})^2 &\\ &&(\frac{y_2}{x_3})^2
		\end{pmatrix},  \h \sum_{i,j,k} A_{ijk}^{2 \dagger}  A_{ijk}^2=\gamma_2^2 \begin{pmatrix}
		(\frac{y_2}{x_1})^2 &&\\ &(\frac{y_3}{x_2})^2 &\\ &&(\frac{y_1}{x_3})^2
	\end{pmatrix},
\end{equation}
which when added together prove the trace-preserving condition (\ref{major}) for the channel $\mathcal{E}$ defined in (\ref{FinalE}).\\

\section{Constructing arbitrary gates}\label{construction}
In the previous section, as a result of majorization condition, we saw that maximally coherent states (\ref{MC1}) or (\ref{MC2}) are the most resourceful states in the context of state conversion.  
Now we also prove that, starting from these states and only by using incoherent operations (IO), one can implement any arbitrary quantum gate $U$.\\

\ni The goal is to perform the unitary operation $ U=\sum_{\mu,\nu=1}^{M} |\mu\rangle \langle \nu | \otimes A_{\mu \nu}  $ on the arbitrary quantum state $|\Psi\rangle= \sum_{\alpha=1}^{M} x_\alpha |\alpha\rangle \otimes |\psi_\alpha \rangle$. Following the same idea as in \cite{Plenio}, and without loss of generality, we use an ancillary system with the maximally coherent state (\ref{MC2}), and we define the joint state $|\xi\ra$,
\be
|\xi \rangle =|\Psi\rangle \bigotimes |\Phi\rangle=|\Psi\ra\bigotimes  
\frac{1}{\sqrt{M}}\begin{pmatrix}|\phi_1\ra\\ |\phi_2\ra\\ \cdot\\  \cdot\\ |\phi_M\ra\end{pmatrix}.
\ee
Now consider the incoherent Kraus operators $\mathcal{K}_s$, $s=1,..,M$, defined as follows 
\begin{equation}
	\mathcal{K}_s=\sum_{\mu,\nu=1}^{M} |\mu\rangle \langle \nu| \otimes A_{\mu\nu}  \bigotimes |s\rangle \langle \mu +s| \otimes  |\phi_s \rangle \langle \phi_{\mu +s}| .
\end{equation}
It is easy to show that $\sum_{s=1}^{M} \mathcal{K}_s^\dagger \mathcal{K}_s=I$, and $\mathcal{K}_s \mathcal{I}_{inc} \mathcal{K}_s^\dagger \subset \mathcal{I}_{inc}$. By straightforward calculations, one finds that 
\be
{\cal K}_s|\xi\ra=\frac{1}{\sqrt{M}}U|\Psi\ra \bigotimes \left( |s\rangle \otimes |\phi_s\ra  \right),
\ee
which leads to the quantum channel
\be
\sum_s {\cal K}_s(|\xi\ra\la \xi|){\cal K}_s^\dagger =U|\Psi\ra\la \Psi|U^\dagger \bigotimes \rho_{inc},
\ee
where 
\be
\rho_{inc}=\frac{1}{M}\sum_\mu  |\mu\ra\la\mu| \otimes |\phi_\mu\ra\la \phi_\mu|,
\ee
is the completely decohered form of the maximally coherent state $|\Phi\ra$ which we started with. Thus by consuming a maximally coherent state we can implement any unitary operator on any arbitrary state.

\section{Block Cohering and decohering power}\label{powers}
Using the definition of block coherence, one can also define the block-cohering and block-decohering powers of a quantum channel $\mathcal{E}$, just like the definitions of \cite{Mani} or \cite{Zanardi} for cohering and decohering powers. Following the definitions presented in \cite{Mani}, the Block-Cohering Power (BCP) and the Block-Decohering Power (BDP) of a channel will be defined respectively as 
\begin{equation}\label{BCP}
	BCP(\mathcal{E})=\max_{\rho_{inc} \in \mathcal{I}_{inc}} C(\mathcal{E}(\rho_{_{inc}})),
\end{equation}
\begin{equation}\label{BDP}
	BDP(\mathcal{E})=\max_{\ket{\Psi}_{MC}}  \left[ C(\ket{\Psi}_{MC}\bra{\Psi})-C(\mathcal{E}(\ket{\Psi}_{MC}\bra{\Psi})) \right],
\end{equation}
where $C$ is any well defined block-coherence measure, $\rho_{inc}$ is chosen from the set of block-incoherent states (\ref{incohstate}), and $\ket{\Psi}_{MC}$ stands for maximally block coherent states of the form (\ref{MC1}). The above equations means that the BCP of a channel is equal to the maximum amount of block coherence that can be generated for an initial block incoherent state, and the BDP of a channel is the maximum amount of block coherence of a maximally block coherent state that is destroyed by the quantum channel.
Using the above definitions, one can now calculate the BCP and BDP of any quantum channel, below we will study some channels that are of practical importance.\\

\subsection{Block Cohering Power}
We first follow the same argument as in \cite{Mani} to show that for any quantum channel ${\cal E}$,  linearity of the channel ${\cal E}$ and convexity of the coherence measure allows us to write 

\be \label{BCP2}
BCP({\cal E})=\max_{|\Psi_{inc}\ra} C({\cal E}(|\Psi_{inc}\ra\la \Psi_{inc}|)),
\ee
where $|\Psi_{inc}\ra$ is an incoherent pure state. Note that an incoherent pure state $|\Psi_{inc}\ra$ has only one non-zero state in a given subspace, i.e.
\be\label{pureincoh}
|\Psi_\nu\ra=|\nu\rangle \otimes |\psi_\nu\rangle  = \begin{pmatrix} 0\\ 0\\ \cdot\\ |\psi_\nu\ra\\ \cdot\\   0\end{pmatrix}.
\ee
We now proceed to show equation (\ref{BCP2}), and then we will prove a theorem and study a few examples.\\

\ni {\bf Lemma:} The BCP of a channel which is defined in (\ref{BCP}) is equal to (\ref{BCP2}).\\

\ni {\bf Proof:} Consider an incoherent state $\rho_{inc}=\sum_{\mu=1}^M p_\mu |\mu\ra\la\mu| \otimes \rho_\mu$, then by considering the pure state decomposition of each $\rho_\mu$, the above incoherent state can be written as 
$$
\rho_{inc}=\sum_{\mu , j} p_\mu q_{\mu}^{(j)} |\mu\ra\la\mu| \otimes |\psi_\mu^{j} \ra \la \psi_\mu^{j}|, 
$$
where $\{q_{\mu}^{(j)} \}$ is a probability distribution for each $\mu$. From there,
\begin{eqnarray}
	C(\mathcal{E}(\rho_{inc}))&=&	C(\sum_{\mu , j} p_\mu q_{\mu}^{(j)} \mathcal{E}(|\mu\ra\la\mu| \otimes |\psi_\mu^{j} \ra \la \psi_\mu^{j}|))\cr &\leq& \sum_{\mu , j} p_\mu q_{\mu}^{(j)} C(\mathcal{E}(|\mu\ra\la\mu| \otimes |\psi_\mu^{j} \ra \la \psi_\mu^{j}|))\cr   &\leq& C(\mathcal{E}( |\alpha \ra\la \alpha | \otimes |\psi_\a^{i} \ra \la \psi_\a^{i}|  )),
\end{eqnarray}
where $\alpha$ and $i$ are the block and state numbers that have the largest value of $C(\mathcal{E}(|\mu\ra\la\mu| \otimes |\psi_\mu^{j} \ra \la \psi_\mu^{j}|))$ among all possible values of $\mu$ and $j$. 
The above equation proves the theorem which states that the maximization of (\ref{BCP}) can only be performed over pure incoherent states.\\

\ni This Lemma not only simplifies the calculation of cohering power, but also gives us an alternative method for characterization of  incoherent Kraus operators. In section (\ref{prelim}), we indicated that any quantum channel whose Kraus operators are of the form (\ref{inco operation}) (exemplified in (\ref{exxx})) cannot produce any coherence.  We now present an alternative proof of this fact. This proof, provides us with tools which enable us to calculate in a direct way the BCP of many other channels.\\
 
\ni {\bf Theorem:} Based on the definition (\ref{BCP}), the Block Cohering Power of any quantum channel whose Kraus operators are of the form (\ref{inco operation}) is zero.  \\

\ni{\bf Proof:}
For definiteness, consider a pure incoherent state of the form $|\Psi_1\ra$ (a similar analysis applies to other states $|\Psi_\mu\ra$).
Consider a quantum channel ${\cal E}$, with Kraus operators of the form $K^i= |\m\ra\la \nu| \otimes K^i_{\mu\nu}$. Note that in each block of $K^i$ we have an operator  $K^i_{\mu\nu}: L(H_\nu)\lo L(H_\mu)$ . The action of this Kraus operator on the staet $|\Psi_1\ra$, leads to a non-normalized vector of the form 
\be
|\Phi^i\ra:=\begin{pmatrix} K^i_{11}|\psi_1\ra\\  K^i_{21}|\psi_1\ra\\ K^i_{31}|\psi_1\ra \\ \cdot \\ \cdot\\ K^i_{M1}|\psi_1\ra \end{pmatrix},
\ee
and from there one can write  
\be
{\cal E}(|\Psi_1\ra\la \Psi_1|)=\sum_{i}|\Phi^i\ra\la \Phi^i|.
\ee
The convexity of coherence measure,  again leads to
\be
C({\cal E}(|\Psi_1\ra\la \Psi_1|))=C(\sum_i(|\Phi^i\ra\la \Phi^i|)) \leq \sum_i C(|\Phi^i\ra\la \Phi^i|),
\ee
and by using (\ref{newcoherence}) for the coherence measure of pure states, we find
\be
C_1({\cal E}(|\Psi_1\ra\la \Psi_1|))\leq \sum_i \sum_{\alpha\ne \beta} \sqrt{\la \psi_1| {K^i_{\beta 1}}^\dagger K^i_{\beta 1} |\psi_1\ra\la \psi_1| {K^i_{\alpha 1}}^\dagger K^i_{\alpha 1} |\psi_1\ra}.
\ee
This means that if a quantum channel is such that all its Kraus operators have only one non-zero element in each column block, then the cohering power of that channel is zero. This is in accord with our previous statement in (\ref{inco operation}). \\

\ni {\bf Example 1: The BCP of a unitary operator}\\

\ni Let ${\cal E}_u(\rho)=U\rho U^\dagger$ be a unitary channel acting on a pure incoherent state $|\Psi_{\nu}\ra= |\nu\rangle \otimes |\psi_\nu \rangle $, where the non-zero state lives in the $\nu$-th block.   
The block structure of $U$ is revealed when we write it as 
$U=\sum_{\a\beta} |\a\ra\la \beta| \otimes A_{\a\beta}$. One then finds 
$U|\Psi_{\nu}\ra=\begin{pmatrix}A_{1\nu}|\psi_\nu\ra\\ A_{2\nu}|\psi_\nu\ra\\  \cdot\\ \cdot \\  A_{M\nu}|\psi_\nu\ra\end{pmatrix}$.
One then finds from (\ref{newcoherence}) and (\ref{BCP2})
\ba
C_1^M(U|\Psi_{\nu}\ra)&=&\sum_{\mu\ne \mu'}\sqrt{\la \psi_\nu|A_{\mu \nu}^\dagger A_{\mu \nu}|\psi_\nu\ra\la \psi_\nu|A_{\mu'\nu }^\dagger A_{\mu' \nu}|\psi_\nu\ra}\cr
&=& \sum_{\mu,\mu'}\sqrt{\la \xi_{\mu,\nu}|\xi_{\mu,\nu}\ra\la \xi_{\mu',\nu}|\xi_{\mu'\nu}\ra}-\sum_\mu \la \xi_{\mu,\nu}|\xi_{\mu,\nu}\ra\cr
&=& \big(\sum_{\mu}\sqrt{\la \xi_{\mu,\nu}|\xi_{\mu,\nu}\ra} \ \big)^2-1,
\ea
where
$|\xi_{\mu,\nu}\ra=A_{\mu \nu}|\psi_\nu\ra$ and in the last line we have used unitarity of $U$ to set $\sum_\mu \la \xi_{\mu,\nu}|\xi_{\mu,\nu}\ra=1.$ Therefore by taking the initial incoherent state to be a state where $|\psi\ra$ can be in any of the rows, we find the BCP of a general unitary operator:
\be\label{BCPU}
BCP(U)=\max_{\{\nu,|\psi_\nu\ra\}} \Big(\sum_{\mu}\sqrt{\la \xi_{\mu\nu}|\xi_{\mu\nu}\ra} \ \Big)^2-1,
\ee
where
\be\label{xi-mu-nu}
|\xi_{\mu\nu}\ra=A_{\mu\nu}|\psi_\nu\ra.
\ee

\ni As the simplest case, let $M=2$ and  consider the cohering power of a unitary operator 
$U=\begin{pmatrix}A&B\\ C&D\end{pmatrix}$,  acting on $\mathcal{H}=H_{d_1}\oplus H_{d_2}$, where $A, B, C$ and $D$ are $d_1 \times d_1$, $d_1 \times d_2$, $d_2\times d_1$ and $d_2 \times d_2$  dimensional respectively.
Following (\ref{BCPU}), we find for this unitary operator
\be
BCP(U)=\max_{|\psi_1\ra,|\psi_2\ra}\{ (\sqrt{\la \psi_1|A^\dagger A|\psi_1\ra}+\sqrt{\la \psi_1|C^\dagger C|\psi_1\ra} \ )^2,(\sqrt{\la \psi_2|B^\dagger B|\psi_2\ra}+\sqrt{\la \psi_2|D^\dagger D|\psi_2\ra}\ )^2\}-1
\ee
As an explicit example, let $M=2$ and $U$ be a unitary operator acting on  $\mathcal{H}=H_{2} \otimes H_N=H_N \oplus H_N$, of the form $U=\begin{pmatrix} a \mathbb{I}_N & b V\\ -b^* V^\dagger & a^* \mathbb{I}_N\end{pmatrix}$, where $a,\ b$ are complex numbers subject to $|a|^2+|b|^2=1$, $\mathbb{I}_N$ is the identity operator and $V$ is an arbitrary unitary operator acting on $H_N$. For this operator, a simple calculation shows that 
\be
BCP(U)=(|a|+|b|)^2-1=2|ab|,
\ee
which shows that the Hadamard-like block operator $U=\frac{1}{\sqrt{2}}\begin{pmatrix}  \mathbb{I}_N &  V\\ - V^\dagger &  \mathbb{I}_N\end{pmatrix}$ has maximum BCP as it should. \\

\ni {\bf Example 2: The BCP of the tensor product of two operators:}\\

\ni Consider the tensor product of two unitary operators $W=U\otimes V$, where $U$ is $M$ dimensional and $V$ is $N$ dimensional. This operator acts on ${\cal H}=\oplus_{\mu=1}^M { H}_\mu$ where all the subspaces ${ H}_\mu$ are $N$ dimensional. Each block $\mu\nu$ of the unitary matrix $W$ is of the form
$W_{\mu\nu}=u_{\mu\nu} V$, where $u_{\mu\nu}$ is a complex number and is the $\mu\nu$-th entry of the unitary matrix $U$. Inserting this in equation (\ref{xi-mu-nu}), we see that $|\xi_{\mu\nu}\ra=u_{\mu\nu}V|\psi_\nu\ra$ and following the result (\ref{BCPU}), we find 
\be
BCP(U\otimes V)=\max_{\nu} \Big(\sum_{\mu}\sqrt{u_{\mu\nu}u^*_{\mu\nu}}\Big)^2-1.
\ee
But this is nothing but the ordinary cohering power of a unitary matrix $U$ as defined in \cite{Mani}. Thus we have shown that understandably the BCP of a unitary operator  $U\otimes V$ is nothing but the ordinary cohering power of the unitary matrix $U$, as the matrix $V$ acts within each block and it is the matrix $U$ which acts between blocks. \\

\ni {\bf Example 3: The BCP of a random unitary channel:}\\

\ni Consider now a random unitary operator of the form $${\cal E}(\rho)=\sum_{i}p_i(U^i\otimes V^i)\rho (U^i \otimes V^i)^\dagger,$$ acting on $\mathcal{H}=({H}_M\otimes { H}_N)$. By considering the block structure $\mathcal{H}=\oplus_{\mu=1}^{M} {H}_\mu$, where all the subspaces ${H}_\mu$ are $N$ dimensional, from (\ref{BCP2}), we have
\be
BCP({\cal E})=\max_{|\Psi_{inc}\ra}C_{1}^M\Big(\sum_ip_i(U^i\otimes V^i)|\Psi_{inc}\ra\la \Psi_{inc}| (U^i)^\dagger \otimes (V^i)^\dagger\Big).
\ee
As in previous examples, consider the pure incoherent input state $|\Psi_\nu\ra$, with the only non zero entity $|\psi_\nu\ra$ in the  $\nu$-th block. One finds after straightforward calculations

\be
{\cal E}(|\Psi_\nu\ra\la\Psi_\nu |)=\begin{pmatrix}
	B^{\nu}_{11} & B^{\nu}_{12} & .& B^{\nu}_{1M}\\
	B^{\nu}_{21} & B^{\nu}_{22} & .& B^{\nu}_{2M}\\
	. &.&.&.\\
	B^{\nu}_{M1} & B^{\nu}_{M2} & .& B^{\nu}_{MM}
\end{pmatrix},
\ee
where $B^{\nu}_{\mu \mu'}$ are the following $N$ dimensional matrices
\begin{equation}
	B^{\nu}_{\mu \mu'}=\sum_{i} p_i (U^i)_{\mu \nu} \overline{(U^{i})_{\mu' \nu}}  \ V_i |\psi_\nu\ra\la\psi_\nu|V_i^\dagger,
\end{equation}
In view of the relations (\ref{l1norm}) and (\ref{BCP2}), we find
\begin{eqnarray}
	BCP(\mathcal{E})&=&\max_{\nu,|\psi_\nu\ra} \sum_{\mu\neq\mu'} \Vert B^\nu_{\mu \mu'} \Vert_1.
\end{eqnarray}

\ni As a very simple example, one finds after some simple calculations that for the channel ${\cal E}(\rho)=(1-p)\rho + (U\otimes V)\rho(U\otimes V)^\dagger$ acting on two qubits, with $U=\begin{pmatrix}a& b \\-b^*&a \end{pmatrix}$, one finds 
$BCP({\cal E})=2p|ab|$.
	
\subsection{Block Decohering Power}	
\ni In this section, we use (\ref{BDP}) and calculate the Block Decohering Power of a few channels. \\

\ni {\bf Example: The BDP of a unitary channel:}\\

\ni Let ${\cal E}_u(\rho)=U\rho U^\dagger$ be a unitary channel acting on a maximally block coherent state $|\Psi_{MC}\ra=\frac{1}{\sqrt{M}}\sum_{\mu}|\mu\ra \otimes |\psi_\mu\ra$. The block coherence of this state is from (\ref{MaxCoh}) equal to $C_1^M(|\Psi_{MC}\ra)=M-1$.  The block structure of $U$ is revealed when we write it as 
$U=\sum_{\a\beta} |\a\ra\la \beta| \otimes A_{\a\beta}$. One then finds 
$U|\Psi_{MC}\ra=\frac{1}{\sqrt{M}}\sum_{\a\mu} |\a\ra \otimes A_{\a\mu}|\psi_\mu\ra$. The block coherence of this state is determined from () to be 
\be
C_1^M(U|\Psi_{MC}\ra)=\frac{1}{M}\sum_{\a\ne \a'}\sqrt{\la \chi_\a|\chi_\a\ra\la\chi_{\a'}|\chi_{\a'} \ra},
\ee
where $|\chi_\a\ra=\sum_\mu A_{\a\mu}|\psi_\mu\ra$. This can be rewritten as
	\be
C_1^M(U|\Psi_{MC}\ra)=\frac{1}{M}\Big[\big(\sum_{\a}\sqrt{\la \chi_\a|\chi_\a\ra}\ \big)^2-\sum_\a \la \chi_\a|\chi_\a\ra	\Big].
	\ee
		Using unitarity of $U$, we note that 
		$
		\sum_\a\la \chi_\a|\chi_\a\ra=\sum_\a\sum_{\mu,\nu}\la \psi_\mu|\psi_\mu\ra=M.
		$
The block decohering power will then be
		\be
		BDP({\cal E}_u)=M-\frac{1}{M} \min_{\{\psi_\mu\}}\big(\sum_\a \sqrt{\la \chi_\a|\chi_\a\ra} \ \big)^2.
		\ee
As an explicit example, let $M=2$ and $U$ be a unitary operator acting on $\mathcal{H}=H_2\otimes {H}_N$ of the form $U=\begin{pmatrix} a I_N & b V\\ -b^* V^\dagger & a^* I_N\end{pmatrix}$, where $a,\ b$ are complex number subject to $|a|^2+|b|^2=1$, $I_N$ is the identity operator and $V$ is an arbitrary unitary operator actiong on ${H}_N$. For this unitary operator we have
\be
|\chi_1\ra=a|\psi_1\ra+bV|\psi_2\ra,\ \ \ \ \ |\chi_2\ra=-b^* V^\dagger|\psi_1\ra+a^*|\psi_2\ra,\ 
\ee
leading to 
\be
BDP(U)=2-\frac{1}{2} \min_{\{|\psi_1\ra,|\psi_2\ra\}}\Big(\sqrt{1+x}+\sqrt{1-x}\Big)^2 , 
\ee		
where $x=2 Re(a^* b \la \psi_1|V|\psi_2\ra)$.
The minimum value of the function $f(x)=\sqrt{1+x}+\sqrt{1-x}$	is obtained at $x=\pm 1$. This demands that the  maximally coherent state in (\ref{BDP}) which defines the decohering power of the above unitary operator $U$ should be chosen such that
\be
|\psi_2\ra=e^{-i arg(a^* b)}V^\dagger |\psi_1\ra, 
\ee
and it leads to the following BDP for the operator $U$, 
\be
BDP(U)=1-\sqrt{1-4|ab|^2}.
\ee		
Understandably for any block-diagonal or block-anti-diagonal operator, this will give zero BDP and for the block Hadamard operator $H=\frac{1}{\sqrt{2}}\begin{pmatrix}  \mathbb{I}_N &  \mathbb{I}_N\\ \mathbb{I}_N & - \mathbb{I}_N\end{pmatrix}$, it will give $BDP({\cal E}_H)=1.$		This last example is in fact a manifestation of a more general pattern which can be proved by simple and similar equation for any block structure. \\

\ni {\bf Proposition: } For any unitary operator $U\otimes V$ acting on  $\mathcal{H}=H_M\otimes {H}_N=\oplus_{\mu=1}^M H_N$, we have $BDP(U\otimes V)=BDP(U)$ which is intuitively plausible. \\

\section{Relation between block-coherence and $k$-coherence} \label{relation}
The original notion of incoherence \cite{Aberg, Plenio} which defines incoherent states as diagonal density matrices in a specific basis, has been aptly generalized to multi-level or $k$-coherence \cite{Kcoherence1, Kcoherence2, Kcoherence3, Kcoherence4 }. In this generalized setting, a state 
\be
|\psi\ra=\sum_{i=1}^d c_i|i\ra,
\ee
is said to have coherence at level $k$, if exactly $k$ of the coefficients $c_i$ are non-zero. Thus an incoherent state has coherence at level $1$, and a state like $|\psi\ra=a|0\ra+b|1\ra$ in $\mathcal{H}_d$ has coherence at level $2$ and so on.  A state with coherence at level $k$, is said to have coherence rank equal to $k$:  
$r_C(|\psi\ra)=k$. The generalization to mixed states is done by defining the states with coherence level $k$ to be the convex combination of all pure states whose coherence level is less than or equal to $k$, i.e. 
\be
\mathcal{C}_k:=conv\{|\psi\ra\la \psi|, \ | \ r_C(|\psi\ra)\leq k\}.
\ee
Obviously these sets obey the following inclusion relation: 
\be
\mathcal{C}_1\subset \mathcal{C}_2\subset \mathcal{C}_3 \subset \cdots \subset \mathcal{C}_d.
\ee

\begin{figure}[t]
	\centering
	\includegraphics[width=15cm]{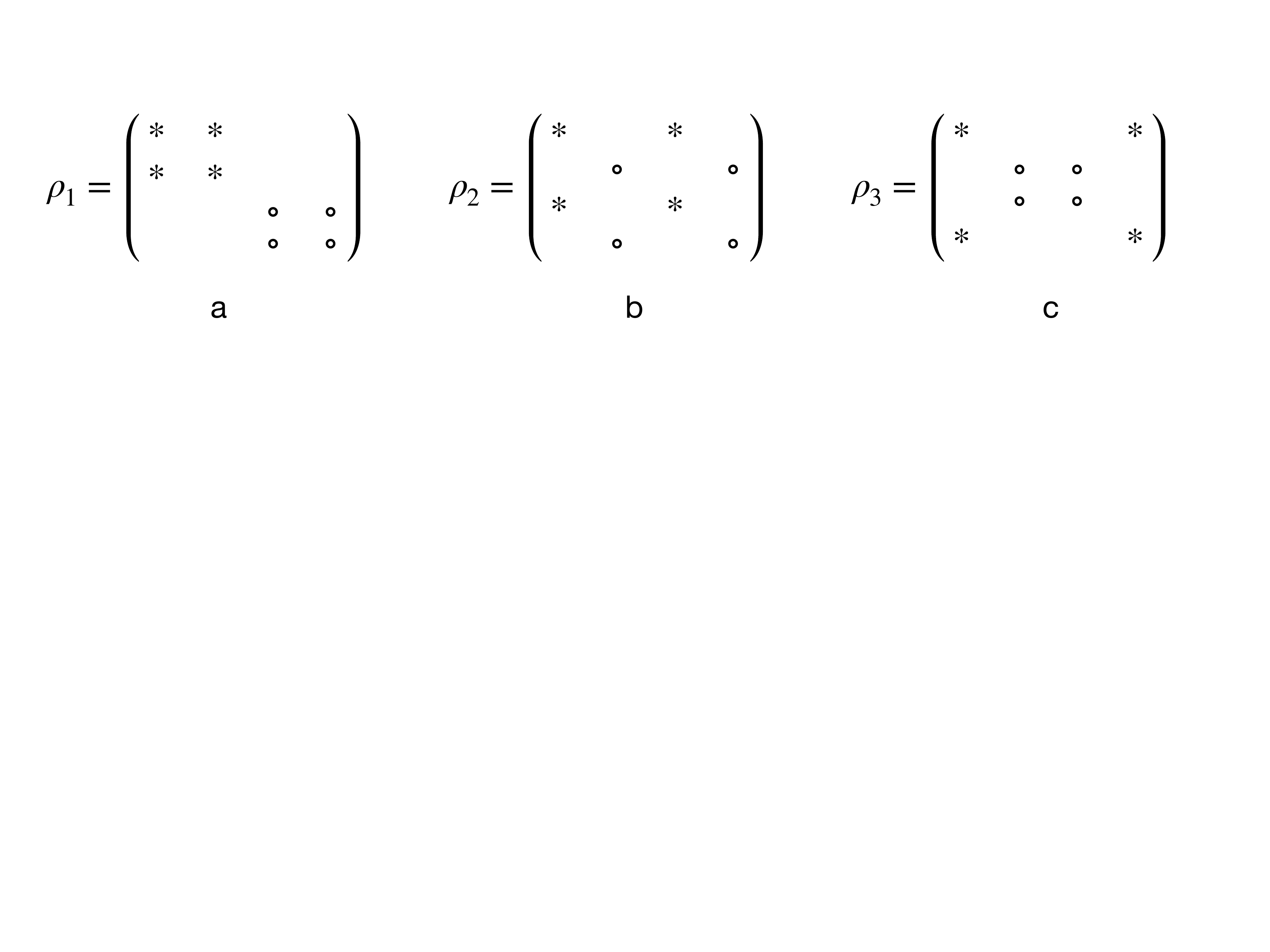}\vspace{-7.5cm}
	\caption{Three different block structures for a density matirx of two particles. The basis states are ordered as $|00\ra, |01\ra, |10\ra$ and $|11\ra$. These correspond to three different block structures $\mathcal{H}_4=H_2\oplus H_2$. All these states belong to $\mathcal{C}_2\subset \mathcal{H}_4$.  }
	\label{k-coherence}
\end{figure}

\ni The relation between $k$-coherence and block coherence is interesting, and we explore it in this section. For the sake of simplicity we describe this relation by presenting an explicit simple example. The basic idea can then be understood in the general setting. Consider a density matrix  $\rho\in L(\mathcal{H}_4)$ .  It is physically more interesting to consider the example of two particles (ions in an ion trap), although this restriction is not necessary. Thus $\mathcal{H}_4$ is the four dimensional space of two qubits and the preferred basis is taken to be $\{|00\ra,|01\ra, |10\ra, |11\ra\}$. Figure (\ref{k-coherence}) shows three different block structures of this matrix.  \\

\ni In figure (\ref{k-coherence}-a)
the density matrix is given by 
\be
\rho^{(1)}=|0\ra\la 0|\otimes \rho_0+|1\ra\la 1|\otimes \rho_1 ,
\ee
which indicates that the first particle has no coherence at all, due to a measurement of the first particle in the computational basis. Decomposition of the states $\rho_0$ and $\rho_1$ casts this state into the form
\be
\rho^{(1)}= \sum_i |\psi_i\ra\la \psi_i|+\sum_i|\phi_i\ra\la \phi_i|,
\ee
where 
\be
|\psi_i\ra=\a_i |0,0\ra+\beta_i|0,1\ra,\ \ \ {\rm and}\ \ \ |\phi_i\ra=\gamma_i |1,0\ra+\delta_i|1,1\ra.
\ee
(Note that in the above equations and in the ones that follow in this section, we use a minimal notation, in order not to clutter the notation. Thus we use non-normalized states and density matrices, and we also use repetitive symbols.)
This shows that $\rho^{(1)}$ is the convex combination of coherent states of level $2$ and thus $\rho^{(1)}\in \mathcal{C}_2$. However not all states of $\mathcal{C}_2$ are of this form, since not all $2$-coherent pure states are involved in this decomposition. Consider now another block structure shown in figure (\ref{k-coherence}-b), induced by measurements on the second particle, again in the computational basis.  Following the same argument as before, the state is now given by 
\be
\rho^{(2)}=\rho_0\otimes  |0\ra\la 0|+\rho_1\otimes |1\ra\la 1|,
\ee
or after decomposition of the states $\rho_0$ and $\rho_1$, 
\be
\rho_2=  \sum_i |\psi_i,\ra\la \psi_i|+\sum_i|\phi_i\ra\la \phi_i|,
\ee
where
\be
|\psi_i\ra=\a_i |0,0\ra+\beta_i|1,0\ra,\ \ \ {\rm and}\ \ \ |\phi_i\ra=\gamma_i |0,1\ra+\delta_i|1,1\ra,
\ee
and thus again, we find $\rho^{(2)}\in \mathcal{C}_2$. The states of the form $\rho^{(1)}$ and $\rho^{(2)}$ do not  still comprise all the states of $\mathcal{C}_2$. This is due to the fact that we have not exhausted all the block structures (i.e. measurements). The last block structure is shown in figure (\ref{k-coherence}-c) and is induced by a  measurement with projectors $\pi_0=|00\ra\la 00|+|11\ra\la 11|$ and $\pi_1=|01\ra\la 01|+|10\ra\la 10|$ (i.e. a measurement which determines the equality or difference of the two qubits). The state is then written as 
\be
\rho^{(3)}=\sum_i |\psi_i\ra\la \psi_i|+\sum_i|\phi_i\ra\la \phi_i|,
\ee
where
\be
|\psi_i\ra=\a_i |0,0\ra+\beta_i|1,1\ra,\ \ \ {\rm and}\ \ \ |\phi_i\ra=\gamma_i |0,1\ra+\delta_i|1,0\ra.
\ee
This shows that $\rho^{(3)}\in \mathcal{C}_2$ too, and any state in $\mathcal{C}_2$ is either of the form $\rho^{(1)}$, $\rho^{(2)}$ or $\rho^{(3)}$. After seeing this simple example, we are ready to state the relation between block coherence and $k$-coherence.\\

\ni Suppose that we have a block structure $B_k$ based on the decomposition of the Hilbert space $\mathcal{H}_d=\oplus_{\mu=1}^M H_\mu$, subject to the following constraint 
\be \label{dim}
dim(H_\mu)\leq k, \ \ \ \ \forall \ \mu.
\ee
Then, according to equation (\ref{incohstate}) the set of incoherent states with regard to block structure $B_k$ is
\begin{equation}
	\mathcal{I}_{inc}^{(B_k)}=\{ \rho \ | \ \rho=\sum_{\mu} \pi_\mu \rho \pi_\mu  \ \  \& \ \ rank(\pi_\mu)\leq k \},
\end{equation}
where $\pi_\mu$ is the projection operator on the subspace $H_\mu$. We now conjecture the following relation between block-coherence and $k$-coherence
\be
\bigcup_{B}  \mathcal{I}_{inc}^{(B_k)}= \mathcal{C}_k,
\ee
where $\cup_B$ means a union over all block structures of the form (\ref{dim}). \\

\ni In passing, one may be tempted to ask why an alternative definition of $k$-coherence has not been adopted from the very beginning for mixed states? i.e. one  in which diagonal density matrices are $1$-coherent states, three-diagonal density matrices are  $2$-coherent states, five-diagonal states  are $2$- coherent states etc? We think that while this categorization is in principle possible, it is not motivated by physical measurements, even on adjacent particles in a many-body system. The simple two-particle system that we have analyzed in this section may not show this clearly, but it is easily seen in a three-particle system with basis states $\{|000\ra, |001\ra, |010\ra, |011\ra, |100\ra, |101\ra, |110\ra, |111\ra\}$, that measurement of the second particle in the basis $\{|0\ra, 1\ra\}$, entails a block structure which contain non-zero elements far from the diagonal. To our understanding, this explains why the definition of $k$-coherence as adopted in \cite{Kcoherence1, Kcoherence2, Kcoherence3, Kcoherence4} is the natural one.  

\section{Conclusion} \label{Conclusion}

The concept of block coherence, based on projective measurement, was first introduced in  \cite{Aberg} and then generalized via Naimark extension in \cite{Bischof1, Bischof2} to include POVM measurements. In these works certain general properties of the resource theory of block coherence were proved.  In the present work, we restrict ourselves to projective measurements and adopt a notational framework, which facilitates many explicit calculations.  In particular, this enables us to prove that a majorization  condition is sufficient and necessary for state transformation using block-incoherent operations  (section \ref{PureStateConversion} and the appendix) .
Moreover, we are able to define the Block-Cohering Power (BCP) and Block-Decohering Power (BDP) of quantum operations (section \ref{powers}), as an extension of the works in \cite{Mani,Zanardi}. \\

\ni This framework makes it  also possible to connect  block-coherence , in a transparent way, with other generalized notions of coherence. An example is the connection with $k$-coherence which is discussed via a simple example in section (\ref{relation}). Within this framework it is also possible to extend other  classes of resource theories to their block form. An example is the Dephasing covariant Incoherent Operations (DIO). In ordinary resource theory of coherence, a quantum operation ${\cal E}$ is a DIO operation, if it commutes with the dephasing operation 
$\Delta: \rho\lo \sum_{\mu} |\mu\ra\la \mu|\rho|\mu\ra\la \mu|$. Many of the results in \cite{Gour1, Gour2, Marvian} on this kind of resource theory can be readily extended after proper modifications by defining Block-DIO operations as those which commute with the Block-dephasing operator
$\Delta^B: \rho\lo \sum_{\mu} |\mu\ra\la \mu|\otimes \rho_{\mu\nu}$, where $\rho_{\mu}$ is now the operator on a block.  Actually such an extension seems to be present also in \cite{Marvian}, where a large system is partitioned into subsystems each carrying out a different representation of the translation symmetry group.

\section{Acknowledgment} \label{ack}
We would like to thank one of the anonymous referees for his or her very valuable comments and suggestions which led to an extensive revision of this paper in many respects.
%%%%%%%%%%%%%%%%%%%%%%%%%%%%%%%%%%%%%%%%%%%%%%%%%

\section*{Appendix: Proof of necessary condition for state transformation}

In the main text, we  constructed a  block-incoherent opeation with specific forms for the Kraus operators which, provided that $\bx\prec\by $,  transforms the state $|\Phi_\bx\ra$ to $|\Phi_\by\ra$.  We now prove the converse statement:  if there is any block-incoherent operation (with any type of incoherent Kraus operators) which transforms $|\Phi_\bx\ra$ into $|\Phi_\by\ra$, then necessarily the majorization condition holds, that is $\bx\prec \by$. Thus majorization is both a necessary and sufficient condition for this transformation. The basic idea of the proof of necessity can be conveyed in the simple case where we have two subspaces, i.e. $M=2$. This saves us and the reader from cluttered formulas and notations. The argument for the general case of arbitrary number of subspaces, is a straightforward generalization. \\

\noindent So let  $\mathcal{H}=H_1\oplus H_2 $, and suppose that there is an incoherent operation $\mathcal{E}(\rho)=\sum_n K_n \rho K_n^\dagger$, which converts the initial state 
\begin{equation}
	|\Phi_{\bx}\ra = \begin{pmatrix} x_1 |\phi_1\rangle \\ x_2 |\phi_2\rangle 
	\end{pmatrix},\h x_1^2+x_2^2=1,
\end{equation}
to the final state 
\begin{equation}
	|\Phi_{\by}\ra = \begin{pmatrix} y_1 |\phi_1\rangle \\ y_2 |\phi_2\rangle 
	\end{pmatrix},\h y_1^2+y_2^2=1,
\end{equation}
where without loss of generality, we have taken the coefficients $x_\mu$ and $y_\mu$ to be real.  We will now prove that if
\begin{equation} \label{map}
	\sum_a K_a |\Phi_\bx\ra\la \Phi_\bx| K_a^\dagger =|\Phi_\by\ra\la \Phi_\by|,
\end{equation}
then $\textbf{x}\prec \textbf{y}$ where $\textbf{x}=(x_1^2 , x_2^2 )$ and $\textbf{y}=(y_1^2 , y_2^2 )$.\\

\noindent According to (\ref{inco operation}), the general form of an incoherent Kraus operator is such that it has exactly only one nonzero block in each column, and can be written in the form of 
\be
K_a = \sum_\mu   |a(\mu)\ra\la \mu| \otimes K^a_\mu,
\ee
in which $a: \{1, 2\}\lo \{1,2\}$ is an arbitrary function. By using a  suitable permutation $P_a$ on the blocks,  the above Kraus operator can be cast into the form  
\be
K_a = P_a \sum_\mu   |a(\mu)\ra\la \mu| \otimes K^a_\mu,
\ee
where $a(\mu)$ is now restricted such that $1\leq a(\mu) \leq \mu$. 
 Hence, without loss of generality, we can write the following form for the incoherent Kraus operator $K_a$;
\begin{equation} 
	K_a= P_a \left(\begin{array}{ccc}
		K_1^a &  \delta_{1,a(2)} K_2^a  \\
		0 & \delta_{2,a(2)} K_2^a 
	\end{array}\right). 
\end{equation}
The permutation matrix $P_a$ preceding the upper triangular Kraus operator effectively covers all the possible forms of the incoherent Kraus operators $K_a$.  From the condition $\sum_a K_a^\dagger K_a =I$, we get 
\begin{eqnarray}\label{ttt}
	\begin{cases}
		\sum_a {K_{\mu}^{a}}^\dagger K_{\mu}^a = I_\mu  \hspace{0.25 cm} \mu=1,2 , \cr
		\sum_a \delta_{1,a(2)} {K_{1}^{a}}^\dagger K_{2}^a= \textbf{0}_{d_1 \times d_2} .
	\end{cases}
\end{eqnarray}

\noindent On the other hand, according to equation (\ref{map}), for each $a$ there exist complex number $\alpha_a$, such that $K_a |\Phi_{\textbf{x}}\rangle =\alpha_a |\Phi_{\textbf{y}}\rangle$, and hence 

\begin{equation}
	P_a \left(\begin{array}{c}
		x_1 K_1^a |\phi_1 \ra +x_2 \delta_{1,a(2)} K_2^a |\phi_2 \ra  \cr
		x_2 \delta_{2,a(2)} K_2^a |\phi_2 \ra \
	\end{array}\right)=
	\alpha_a \left(\begin{array}{c}
		y_1 |\phi_1 \ra \cr y_2 |\phi_1 \ra 
	\end{array}\right),
\end{equation}
or equivalently
\begin{equation}\label{bbb}
	\left(\begin{array}{c}
		x_1 K_1^a |\phi_1 \ra +x_2 \delta_{1,a(2)} K_2^a |\phi_2 \ra  \cr
		x_2 \delta_{2,a(2)} K_2^a |\phi_2 \ra 
		\end{array}\right)=
	\alpha_a \left(\begin{array}{c}
		y_{P_a^{-1}(1)} |\phi_{P_a^{-1}(1)} \ra \cr y_{P_a^{-1}(2)} |\phi_{P_a^{-1}(2)} \ra 
	\end{array}\right),
\end{equation}
where $P_a^{-1}$ is the inverse of the permutation operator $P_a$.\\

\noindent Equating the norms of vectors in each block on both sides of (\ref{bbb})  and summing over $a$ and using (\ref{ttt}), we find
\begin{eqnarray}
	\begin{cases}
		x_1^2 +  x_2^2 \sum_a \delta_{1,a(2)}  = \sum_a |\alpha_a|^2  y_{P_a^{-1}(1)}^2 ,  \cr
		x_2^2 \sum_a \delta_{2,a(2)}= \sum_a |\alpha_a|^2   y_{P_a^{-1}(2)}^2  .
			\end{cases}
\end{eqnarray}
From the first equation, it is evident that
\begin{equation}
	x_1^2 \leq \sum_a |\alpha_a|^2 y_{P_a^{-1}(1)}^2,
\end{equation}
By adding first and second equations, one will also find that
\begin{equation}
	x_1^2 + x_2^2 = \sum_a |\alpha_a|^2 y_{P_a^{-1}(1)}^2+ \sum_a |\alpha_a|^2 y_{P_a^{-1}(2)}^2.
\end{equation}
From the above two equations, it is evident that 
\begin{equation} \label{part1}
	(x_1^2,x_2^2)\prec (\sum_a |\alpha_a|^2 y_{P_a^{-1}(1)}^2 , \sum_a |\alpha_a|^2 y_{P_a^{-1}(2)}^2). 
\end{equation}

\noindent Now note that for $\mu=1,2$, 
\begin{equation}
	\sum_a |\alpha_a|^2 y_{P_a^{-1}(\mu)}^2 =\sum_{a , P_a^{-1}(\mu)=1} |\alpha_a|^2  \  y_1^2+ \sum_{a,P_a^{-1}(\mu)=2} |\alpha_a|^2  \ y_2^2.
\end{equation}
Let $b_{\mu \nu}:=\sum_{a,P_a^{-1}(\mu)=\nu} |\alpha_a|^2$, for $\mu,\nu\in \{1,2\}$, then, in view of the relation $\sum_a |\alpha_a|^2 =1$,  the matrix $B=(b_{\mu \nu})$ is a doubly stochastic matrix, and
\begin{equation}
	B(y_1^2 , y_2^2 )^t =(\sum_a |\alpha_a|^2 y_{P_a^{-1}(1)}^2, \sum_a |\alpha_a|^2 y_{P_a^{-1}(2)}^2)^t,
\end{equation}
which implies that \cite{nielson}
\begin{equation} \label{part2}
	(\sum_a |\alpha_a|^2 y_{P_a^{-1}(1)}^2, \sum_a |\alpha_a|^2 y_{P_a^{-1}(2)}^2) \prec (y_1^2 , y_2^2 ).
\end{equation}

\noindent From the equations (\ref{part1}) and (\ref{part2}), one infers that
\begin{equation}
	(x_1^2 , x_2^2 )\prec (y_1^2 , y_2^2).
\end{equation}
This proves the theorem. 

\begin{thebibliography}{}

% \bibitem{coherence application1} N. Gisin, G. Ribordy, W. Tittel, and H. Zbinden, \textit{Quantum cryptography}, Rev. Mod. Phys. \textbf{74}, 145–195 (2002).
%“Quantum cryptography”

%\bibitem{coherence application2} V. Giovannetti, S. Lloyd, and L. Maccone,  Nature Photonics \textbf{5}, 222 (2011).
%“Advances in quantum metrology”

%\bibitem{coherence application3} M. A. Nielsen and I. L. Chuang, Quantum Computation and Quantum Information. Cambridge University Press (2010).

%\bibitem{coherence application4} N. Gisin and R. Thew,  Nature Photonics \textbf{1}, 165 (2007).
%“Quantum communication”
\bibitem{Aberg} J. Åberg, \textit{Quantifying Superposition}, arXiv:quant-ph/0612146, (2006).

\bibitem{Bischof1} F. Bischof, H. Kampermann, and D. Bruß, \textit{Resource theory of coherence based on positive-operator-valued measures}, Phys. Rev. Lett. \textbf{123}, 110402 (2019).

\bibitem{Bischof2} F. Bischof, H. Kampermann, and D. Bruß, \textit{Quantifying coherence with respect to general quantum measurements}, Phys. Rev. A. \textbf{103}, 032429 (2021).


\bibitem{marcus1}
M. Arndt, O. Nairz, J. Vos-Andreae, C. Keller, G. Van der Zouw, and A. Zeilinger, \textit{Wave–particle duality of C60 molecules}, Nature \textbf{401}, 680-682 (1999).

\bibitem{marcus2}
S. Gerlich, S. Eibenberger, M. Tomandl, and et al., \textit{Quantum interference of large organic molecules},Nature communications \textbf{2}, 263 (2011).

\bibitem{fazio1}
L. Amico, R. Fazio, A. Osterloh, and V. Vedral, \textit{Entanglement in many-body systems}, 
Rev. mod. phys. \textbf{80}, 517 (2008).

\bibitem{tahereh}
T. Abad, and V. Karimipour, \textit{Scaling of macroscopic superpositions close to a quantum phase transition}, Phys. Rev. B \textbf{93}, 195127 (2016).

\bibitem{nielson}
M. A. Nielson and I. L. Chuang, \textit{Quantum Computation and Quantum Information}, Cambrdige University Press (2000).

\bibitem{Glauber} R.J. Glauber, \textit{Coherent and Incoherent States of the Radiation Field}, Phys. Rev. \textbf{131}, 2766 (1963).

\bibitem{Sudarshan} E. C. G. Sudarshan, \textit{Equivalence of Semiclassical and Quantum Mechanical Descriptions of Statistical Light Beams}, Phys. Rev. Lett. \textbf{10}, 277 (1963).

\bibitem{Mandel} L. Mandel, and E. Wolf, \textit{Coherence Properties of Optical Fields}, Rev. Mod. Phys. \textbf{37}, 231 (1965).

\bibitem{sup1} M. Oszmaniec, A. Grudka, M. Horodecki, and A. Wojcik, \textit{Creation of superposition of unknown quantum states} , Phys. Rev. Lett. \textbf{116}, 110403 (2016).

\bibitem{sup2} M. Doosti, F. Kianvash, and V. Karimipour, \textit{Universal superposition of orthogonal states}, Phys. Rev. A, \textbf{96}, 052318 (2017).


\bibitem{Plenio} T. Baumgratz, M. Cramer, and M. B. Plenio, \textit{Quantifying Coherence}, Phys. Rev. Lett. \textbf{113}, 140401 (2014).
%

\bibitem{Levi} F. Levi, and F. Mintert, \textit{A quantitative theory of coherent delocalization}, New J. Phys. \textbf{16}, 033007 (2014).
% 






%

% 

%
\bibitem{Kcoherence1} M. Ringbauer, T. R. Bromley, M. Cianciaruso, L. Lami, W. Y. S. Lau, G. Adesso, A. G. White, A. Fedrizzi, and M. Piani, \textit{Certification and Quantification of Multilevel Quantum Coherence}, Phys. Rev. X \textbf{8}, 041007 (2018).

%

\bibitem{Kcoherence2}  N. Johnston, C. K. Li, S. Plosker, Y. T. Poon, and B. Regula,  \textit{Evaluating the robustness of k-coherence and k-entanglement}
Phys. Rev. A \textbf{98}, 022328 (2018).
%

\bibitem{Kcoherence3} J. Sperling,  and  W. Vogel,   \textit{Convex ordering and quantification of quantumness}. Phys. Scr. \textbf{90}, 074024 (2015).
%

\bibitem{Kcoherence4} N. Killoran, F. E. Steinhoff,  and M. B. Plenio, \textit{Converting non-classicality into entanglement.} Phy. Rev. Lett. \textbf{116}, 080402 (2016).





\bibitem{Xu} Jianwei Xu, Lian-He Shao, and Shao-Ming Fei, \textit{Coherence measures with respect to general quantum measurements}
Phys. Rev. A \textbf{102}, 012411 (2020).


% 
\bibitem{blockcoherence measure} L. Fu, F. Yan and T. Gao, \textit{Block-coherence measures and coherence measures based on positive-operator-valued measures}, Communications in Theoretical Physics, \textbf{74}, 025104 (2022).

%



%
\bibitem{Plenio2} T. Theurer, N. Killoran, D. Egloff, and M. B. Plenio, \textit{A resource theory of superposition}, Phys. Rev. Lett. \textbf{119}, 230401 (2017).
% 
\bibitem{Suman} S. Mandal, M. Narozniak, C. Radhakrishnan, Z. Q. Jiao, X. M. Jin, and T. Byrnes, \textit{Characterizing coherence with quantum observables}, Phys. Rev. Research \textbf{2}, 013157 (2020).


\bibitem{Mani} A. Mani, and V. Karimipour, \textit{Cohering and decohering power of quantum channels}, Phys. Rev. A \textbf{92}, 032331 (2015).
% 
\bibitem{Zanardi}  P. Zanardi, G. Styliaris, and L. C. Venuti,  \textit{Coherence-generating power of quantum unitary maps and beyond}, Phys. Rev. A \textbf{95}, 052306 (2017).
% 


\bibitem{Marvian} I. Marvian, and R. W. Spekkens, \textit{How to quantify coherence: Distinguishing speakable and unspeakable notions}, Phys. Rev. A \textbf{94}, 052324 (2016).
% 

\bibitem{Streltsov1} A. Streltsov, U. Singh, H. S. Dhar, M. N. Bera, and G. Adesso, \textit{Measuring Quantum Coherence with Entanglement}, Phys. Rev. Lett. \textbf{115}, 020403 (2015).
% 

\bibitem{Streltsov2} A. Streltsov, S. Rana, P. Boes, and J. Eisert, \textit{Structure of the Resource Theory of Quantum Coherence}, Phys. Rev. Lett. \textbf{119}, 140402 (2017).
%

\bibitem{Winter} A. Winter, and D. Yang, \textit{Operational Resource Theory of Coherence}, Phys. Rev. Lett. \textbf{116}, 120404 (2016).
% 

\bibitem{Gour1} E. Chitambar, and G. Gour, \textit{Critical Examination of Incoherent Operations and a Physically Consistent Resource Theory of Quantum Coherence}, Phys. Rev. Lett. \textbf{117}, 030401 (2016).
% 

\bibitem{Gour2} E. Chitambar, and G. Gour, \textit{Comparison of incoherent operations and measures of coherence}, Phys. Rev. A \textbf{94}, 052336 (2016).
% 


\bibitem{Yadin} B. Yadin, J. Ma, D. Girolami, M. Gu, and V. Vedral, \textit{Quantum Processes Which Do Not Use Coherence}, Phys. Rev. X \textbf{6}, 041028 (2016).

\bibitem{Bromley} T. R. Bromley, M. Cianciaruso, and G. Adesso, \textit{Frozen Quantum Coherence}, Phys. Rev. Lett. \textbf{114}, 210401 (2015).

\bibitem{Adesso} A. Streltsov, G. Adesso, and M. B. Plenio, \textit{Colloquium: Quantum Coherence as a Resource},  Rev. Mod. Phys. \textbf{89}, 041003 (2017).
% 

\bibitem{Kim} S. Kim, C. Xiong, A. Kumar, and J. Wu, \textit{Converting coherence based on positive-operator-valued measures into entanglement}, Phys. Rev. A \textbf{103}, 052418 (2021). 
% 






%
\bibitem{resource theory} E. Chitambar and G. Gour, \textit{Quantum resource theories}, Reviews of modern physics, \textbf{91},025001, (2019).


\bibitem{entresource1} R. Horodecki, P. Horodecki, M. Horodecki, and K. Horodecki, \textit{Quantum entanglement}, Rev. Mod. Phys. \textbf{81}, 865 (2009).

\bibitem{entresource2} M. B. Plenio, and S. Virmani, \textit{An introduction to entanglement measures}, Quantum Inf. Comput. \textbf{7}, 1 (2007).

\bibitem{entresource3} V. Vedral, M. B. Plenio, M. A. Rippin, and P. L. Knight, \textit{Quantifying Entanglement}, Phys. Rev. Lett. \textbf{78}, 2275 (1997).


\bibitem{assymetryResource1} J. A. Vaccaro, F. Anselmi, H. M. Wiseman, and K. Jacobs, \textit{Tradeoff between extractable mechanical work, accessible entanglement, and ability to act as a reference system, under arbitrary superselection rules}, Phys. Rev. A \textbf{77}, 032114 (2008).

\bibitem{assymetryResource2} G. Gour, and R. W. Spekkens, \textit{The resource theory of quantum reference frames: manipulations and monotones}, New J. Phys. \textbf{10} (3), 033023 (2008).

\bibitem{assymetryResource3} G. Gour, I. Marvian, and R. W. Spekkens, \textit{Measuring the quality of a quantum reference frame: The relative entropy of frameness}, Phys. Rev. A \textbf{80},
012307 (2009).

\bibitem{resource1} M. Horodecki, and J. Oppenheim, \textit{(Quantumness in the context of) Resource Theories}, Int. J. Mod. Phys. B \textbf{27}, 1345019 (2013).
%

\bibitem{resource2} L. del Rio, L. Kraemer, and R. Renner, \textit{Resource theories of knowledge}, arXiv:1511.08818 (2015).
%	

\bibitem{resource3} B. Coecke, T. Fritz, and R. W. Spekkens, \textit{ A mathematical theory of resources}, Information and Computation \textbf{250}, 59 (2016).
% 

\bibitem{transform} S. Du, Z. Bai, and Y. Guo, \textit{Conditions for coherence transformations under incoherent operations}, Phys. Rev. A \textbf{91}, 052120 (2015); Erratum Phys. Rev. A \textbf{95}, 029901 (2017).

% 


\bibitem{Gisin} N. Gisin and S. Popescu, \textit{Spin Flips and Quantum Information for Antiparallel Spins},
Phys. Rev. Lett. \textbf{83}, 432 (1999).
% 

\bibitem{Massar} S. Massar and S. Popescu, \textit{Optimal extraction of information from finite quantum ensembles}, Phys. Rev. Lett. \textbf{74}, 1259 (1995).
% 

\bibitem{Bartlett1} S. D. Bartlett, T. Rudolph, and R. W. Spekkens, \textit{Optimal measurements for relative quantum information}, Phys. Rev. A \textbf{70}, 032321 (2004).
% 

\bibitem{Bartlett2}  S. D. Bartlett, T. Rudolph, and R. W. Spekkens, \textit{Reference frames, superselection rules, and quantum information}, Rev. Mod. phys. \textbf{79}, 555 (2007).
% 

\bibitem{ours1} F. Rezazadeh, A. Mani, and V. Karimipour, \textit{Secure alignment of coordinate systems using quantum correlation}, Phys. Rev. A \textbf{96}, 022310 (2017).
% 

\bibitem{ours2} F. Rezazadeh, A. Mani, and V. Karimipour, \textit{Power of a shared singlet state in comparison to a shared reference frame}, Phys. Rev. A \textbf{100}, 022329 (2019).
% 

\bibitem{ours3} F. Rezazadeh, A. Mani, and V. Karimipour, \textit{Quantum key distribution with no shared reference frame}, Quantum Inf. Process \textbf{19}, 54 (2020).
% 

\bibitem{ours4} F. Rezazadeh, and A. Mani, \textit{Encoding the information in relative parameters}, Phys. Lett. A \textbf{407}, 127454 (2021).
% 

\bibitem{hetero} J. C. J. Egues, and J. W. Wilkins, \textit{Spin-dependent phenomena in digital-magnetic heterostructures: Clustering and phase-space filling effects}, Phys. Rev. B \textbf{58}, 24 (1998).
% 

\bibitem{proof} {Proposition (12.11) of \cite{nielson}:} For two probability vectors $\textbf{x}$ and $\textbf{y}$, $\textbf{y} \succ \textbf{x}$ if and only if $\textbf{x}=\sum_j p_j P_j \textbf{y}$ for some probability distribution $p_j$ and permutation matrices $P_j$.


\end{thebibliography}
\end{document}